\documentclass[aps,prl,reprint,superscriptaddress]{revtex4-1}

\usepackage{graphicx}
\usepackage{dcolumn}
\usepackage{bm}
\usepackage{amsfonts}
\usepackage{amsmath}
\usepackage{amssymb}
\usepackage{subfigure}
\usepackage{CJK}
\usepackage{mathptmx}
\usepackage[breaklinks,colorlinks,linkcolor=blue,citecolor=blue,urlcolor=black,bookmarks=false]{hyperref}
\usepackage{placeins}
\usepackage{color}     
\usepackage{upgreek} 
\usepackage{pdfpages}
\usepackage{pgffor}

\def\unit#1{\mathord{\thinspace\rm #1}}

\makeatletter
\AtBeginDocument{\let\LS@rot\@undefined}
\makeatother

\begin{document}
\begin{CJK*}{Bg5}{bsmi}

\title{Commensurability Oscillations in One-Dimensional Graphene Superlattices}%

\author{Martin Drienovsky}%
\affiliation{Institute of Experimental and Applied Physics, University of Regensburg, D-93040 Regensburg, Germany}%
\author{Jonas Joachimsmeyer}
\affiliation{Institute of Experimental and Applied Physics, University of Regensburg, D-93040 Regensburg, Germany}%
\author{Andreas Sandner}
\affiliation{Institute of Experimental and Applied Physics, University of Regensburg, D-93040 Regensburg, Germany}%
\author{Ming-Hao Liu (¼B©ú»¨)}\email{minghao.liu@phys.ncku.edu.tw}
\affiliation{Department of Physics, National Cheng Kung University, Tainan 70101, Taiwan}%
\affiliation{Institute of Theoretical Physics, University of Regensburg, D-93040 Regensburg, Germany}%
\author{Takashi Taniguchi}
\affiliation{National Institute for Materials Science, 1-1 Namiki, Tsukuba 305-0044, Japan}%
\author{Kenji Watanabe}
\affiliation{National Institute for Materials Science, 1-1 Namiki, Tsukuba 305-0044, Japan}%
\author{Klaus Richter}
\affiliation{Institute of Theoretical Physics, University of Regensburg, D-93040 Regensburg, Germany}%
\author{Dieter Weiss}
\affiliation{Institute of Experimental and Applied Physics, University of Regensburg, D-93040 Regensburg, Germany}%
\author{Jonathan Eroms}%
\email{jonathan.eroms@ur.de}
\affiliation{Institute of Experimental and Applied Physics, University of Regensburg, D-93040 Regensburg, Germany}%


\date{\today}

\begin{abstract}
We report the experimental observation of commensurability oscillations (COs) in 1D graphene superlattices. The widely tunable periodic potential modulation in hBN encapsulated graphene is generated via the interplay of nanopatterned few layer graphene acting as a local bottom gate and a global Si back gate. The longitudinal magneto-resistance shows pronounced COs, when the sample is tuned into the unipolar transport regime. We observe up to six CO minima, providing evidence for a long mean free path despite the potential modulation. Comparison to existing theories shows that small angle scattering is dominant in hBN/graphene/hBN heterostructures. We observe robust COs persisting to temperature exceeding $T=150$ K. At high temperatures, we find deviations from the predicted $T$-dependence, which we ascribe to electron-electron scattering.
\end{abstract}

\maketitle

\end{CJK*}

Due to its high intrinsic mobility \cite{Morozov2008}, graphene is an ideal material for exploring ballistic phenomena. Both suspended graphene \cite{Du2008,Bolotin2008} and graphene-hexagonal boron nitride (hBN) heterostructures \cite{dean2010hbntransfer,wang2013oneDcontacts} were employed to demonstrate integer and fractional quantum Hall effects \cite{young2012qhfm,Bolotin2009,Du2009,LiScience2017}, conductance quantization \cite{Tombros2011,Terres2016}, cyclotron orbits \cite{Taychatanapat2013,Lee2016,sandner2015ballistic,yagi2015antidots}, and ballistic effects at p-n-junctions \cite{rickhaus2013ballistic,rickhaus2015guiding,Lee2015}, all requiring
high mobility. 

In particular, several fascinating observations have been made in moir\'{e} superlattices in graphene/hBN heterostructures. In addition to magnetotransport signatures \cite{dean2013hofstadter,hunt2013massiveDF_Hofstadter,ponomarenko2013fractal} of the fractal energy spectrum predicted by Hofstadter \cite{hofstadter1976energy}, Krishna Kumar {\em et al.} recently reported robust $1/B$ periodic oscillations persisting to above room temperature \cite{KrishnaKumar2017}. Those oscillations were ascribed to band conductivity in superlattice-induced minibands, where the group velocity in those minibands enters into the magnetoconductance. Here, the oscillation period is independent of the carrier density and set only by the lattice spacing $a$ via $\Phi_0/\Phi$, where $\Phi_0$ is the magnetic flux quantum and $\Phi$ the flux through one superlattice unit cell.

However, the archetypal effect where a superlattice potential leads to magnetoresistance oscillations due to miniband conductivity, namely Weiss or commensurability oscillations (CO) \cite{weiss1989magnetoresistance}, has not yet been demonstrated in graphene, owing to the challenging task of combining high mobility graphene and a weak nanometer scale periodic potential. Those oscillations arise due to the interplay between the cyclotron orbits of electrons in a high magnetic field and the superlattice potential. For a 1D modulation, pronounced $1/B$-periodic oscillations in the magnetoresistance $R_{xx}$ are observed, with minima appearing whenever the cyclotron diameter $2r_C$ is a multiple of the lattice period $a$, following the relation 
\begin{equation}
2r_C = \left(\lambda-\frac{1}{4}\right)a.
\label{eq:1}
\end{equation}
with $\lambda$ being an integer. This intuitive picture was confirmed in Beenakker's semiclassical treatment \cite{beenakker1989guidingcenterdrift}. Quantum mechanically, without modulation and at high magnetic field, Landau levels are highly degenerate in the quantum number $k_y$. The superlattice potential lifts the degeneracy and introduces a miniband dispersion $E_N(k_y)$ into each Landau level. The miniband width oscillates with both $1/B$ and energy, and flat bands appear whenever Eq.~\eqref{eq:1} is fulfilled. Therefore, the group velocity $v_g = \partial E_N / \partial k_y\,(1/\hbar)$ also oscillates, leading to magnetoconductance oscillations \cite{gerhardts1989novel,winkler1989landau_bands}. Those oscillations persist to higher temperatures than Shubnikov de Haas oscillations (SdHO) since the band conductivity survives thermal broadening of the density of states. 
In contrast to the oscillations in Ref.~\cite{KrishnaKumar2017}, the COs depend on the electron density $n$ through $r_C=\hbar\sqrt{4\pi n/(g_sg_v)}/eB$, where $g_s$ ($g_v$) is the spin (valley) degeneracy.
The commensurability condition Eq.~\eqref{eq:1} also holds in the case of graphene \cite{matulis2007appearance, tahir2007weiss}. What is different, though, is the Landau level spectrum which is equidistant in the case of a conventional 2DEG but has a square root dependence in case of the Dirac fermions in graphene \cite{novoselov2005,zhang2005}. This has been predicted to also modify the COs \cite{matulis2007appearance, tahir2007weiss,nasir2010magnetotransport}. Notably, Matulis and Peeters calculated very robust COs in the quasiclassical region of small fields that should persist up to high temperatures \cite{matulis2007appearance}.

\begin{figure}[t]
	\includegraphics[width=1.00\columnwidth]{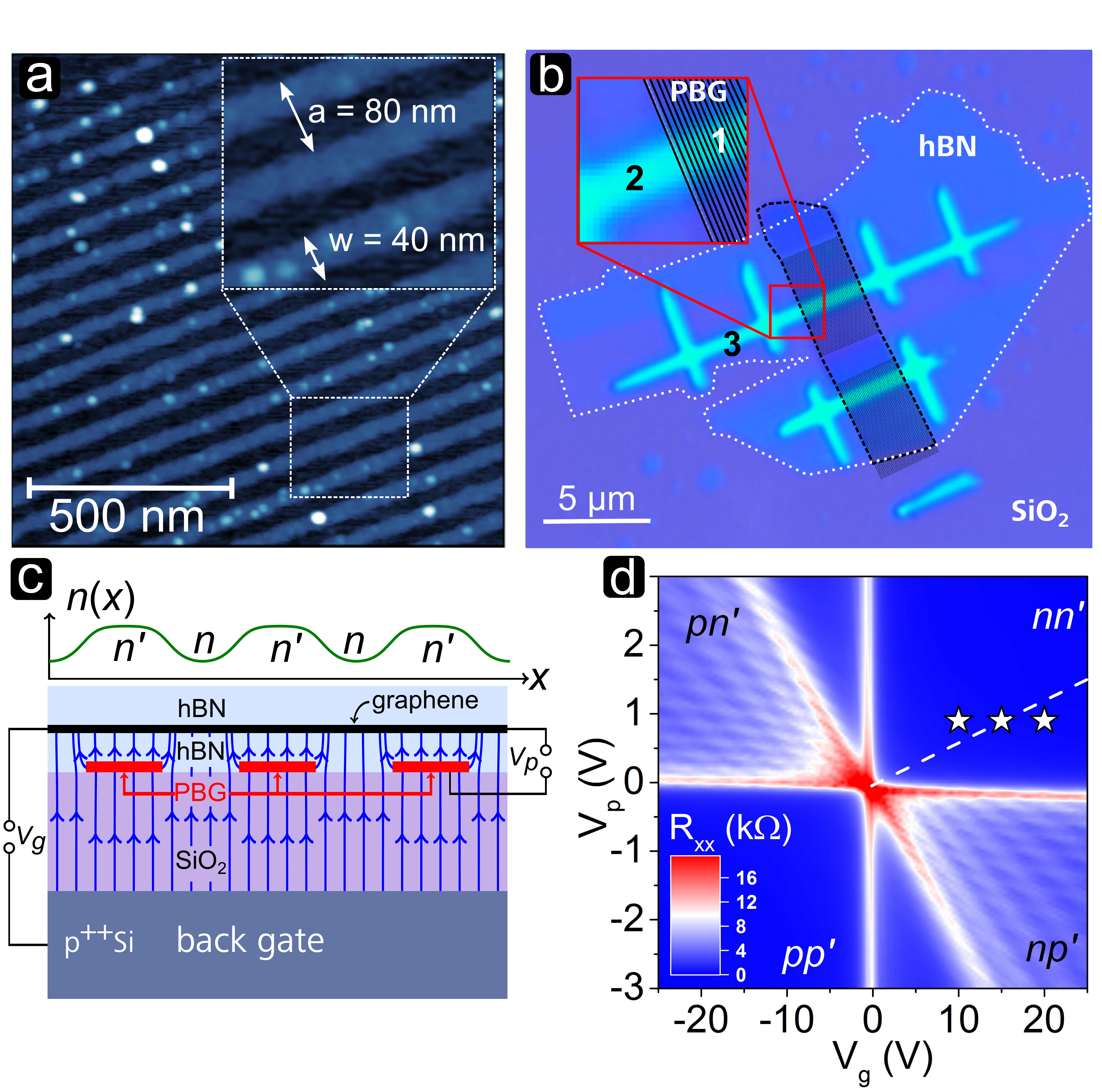}
	\caption{\textbf{Sample geometry and characteristics.} \textbf{a)} AFM image of the PBG of sample B. \textbf{b)} hBN/graphene/hBN heterostructure on top of a few layer graphene patterned bottom gate after mesa etching, before contact deposition (Sample B, PBG outlined in black, lower hBN outlined in white). 
		Labels 1, 2 and 3 denote the modulated, the unmodulated and the reference area, respectively, see text. \textbf{c)} Schematic longitudinal section of the sample geometry, showing the influence of the two independent gates on the graphene charge carrier density in the case of a unipolar modulation. \textbf{d)} Resistance map of sample A  as a function of gate voltages, $V_p$ (PBG) and $V_g$ (backgate) at $B=0$. The highly regular Fabry-P\'erot pattern in the bipolar regions confirms the presence of identical barriers, forming a superlattice. The white dashed line in the $nn^\prime$-quadrant represents the $n=n^\prime$-configuration of the two independent gates. The configurations, marked by stars, will be addressed in the text and Fig. \ref{pic:fig2}.} 	
	\label{pic:fig1}
\end{figure}

Here we employ a patterned few-layer graphene backgating scheme \cite{Li2016,drienovsky2017PBGs} to demonstrate clear cut COs of both Dirac electrons and holes in high-mobility graphene, subjected to a weak unidirectional periodic potential. 
Contrary to hBN/graphene moir\'e lattices, where lattice parameters are set by the materials properties of graphene and hBN, this enables us to define an arbitrary superlattice geometry and strength. As the usual technique of placing a metallic grating with nanoscale periodicity fails due to the poor adhesion of metal to the atomically smooth and inert hBN surface we resort to including a patterned bottom gate (PBG) consisting of few layer graphene 
(FLG) carrying the desired superlattice pattern into the usual van der Waals stacking and edge-contacting technique \cite{wang2013oneDcontacts}. The hBN/graphene/hBN stack is assembled on top of the PBG. Importantly, the bottom hBN layer has to be kept very thin ($<15\unit{nm}$) to impose the periodic potential effectively onto the unpatterned graphene sheet. For the PBG, we exfoliated a FLG sheet (3-4 layers) onto an oxidized, highly p-doped silicon wafer which served as a uniform global back gate in the measurements. The FLG sheet was patterned into the desired shape by electron beam lithography and oxygen plasma etching. This approach exploits the atomic flatness of FLG, which makes it a perfect gate electrode for 2D-material heterostructures 
that can be easily etched into various shapes, e.g.  1D or 2D superlattices, split gates, collimators \cite{cheianov2006selective} or lenses \cite{liu2017lensing}, and allows for nanoscale manipulation of the carrier density. 
Figure \ref{pic:fig1}\textbf{a} shows the AFM image of an $80\unit{nm}$-stripe lattice used for fabrication of sample B, discussed below. The hBN/graphene/hBN stack was deposited onto the PBG, and a mesa was defined by reactive ion etching (Fig. \ref{pic:fig1}\textbf{b}). We used a sequential etching method, employing SF$_6$ \cite{pizzocchero2016hot}, O$_2$ and CHF$_3$/O$_2$-processes, in order to avoid damage to the thin hBN bottom layer covering the PBG (see Supplemental Material for details \cite{Note2}). Edge contacts of evaporated Cr/Au (1 nm/ 90 nm) were deposited after reactive ion etching of the contact region and a brief exposure to oxygen plasma. More details on the fabrication are reported elsewhere \cite{drienovsky2017PBGs}.

The combined action of PBG and the global gate is sketched in Fig. \ref{pic:fig1}\textbf{c}. The PBG partially screens the electric field lines emerging from the Si back gate. The latter therefore controls the carrier type and density in the regions between stripes (labeled $n$), whereas the PBG itself controls primarily those directly above the stripes (labeled $n^\prime$). A typical charge carrier density profile for a weak potential modulation in the unipolar transport regime is shown atop. Hence, tuning both gates separately we can generate unipolar or bipolar potential modulation on the nanoscale.

Transport measurements were performed in a helium cryostat at temperatures between $1.4\unit{K}$ and $200\unit{K}$ and in perpendicular magnetic fields between $0$ to $10\unit{T}$ using low frequency lock-in techniques at a bias current of 10 nA. We present data from two samples (A and B) with a 1D-superlattice period of $a_A= 200\unit{nm}$ and $a_B= 80\unit{nm}$, respectively. The PBGs of both samples, A and B, consist of 19 and 40 stripes of few layer graphene (thickness 3 to 4 layers), respectively. The thicknesses of the lower hBN, separating the graphene from the PBG are $t_A = 13\unit{nm}$ and $t_B =2\unit{nm}$, respectively, measured with AFM. 
Fig.~\ref{pic:fig1}\textbf{d} displays the zero field resistance of sample A. Using both gates, we can tune into the unipolar regime of comparatively low resistance (labeled $nn^\prime$ and $pp^\prime$)  as well as into the bipolar regime (labeled $pn^\prime$ and $np^\prime$), where pronounced Fabry-P\'erot oscillations appear \cite{young2009quantum,rickhaus2013ballistic,drienovsky2014multibarriers,handschin2017fabryperotmoire,drienovsky2017PBGs,dubey2013tunable}. Their regular shape proves the high quality and uniformity of the superlattice potential. The electrostatics of dual gated samples and different transport regimes were discussed, {\em e.g.}, in Refs. \cite{rickhaus2013ballistic,drienovsky2014multibarriers,drienovsky2017PBGs,dubey2013tunable}.

\begin{figure}[t]
	\includegraphics[width=1.05\columnwidth]{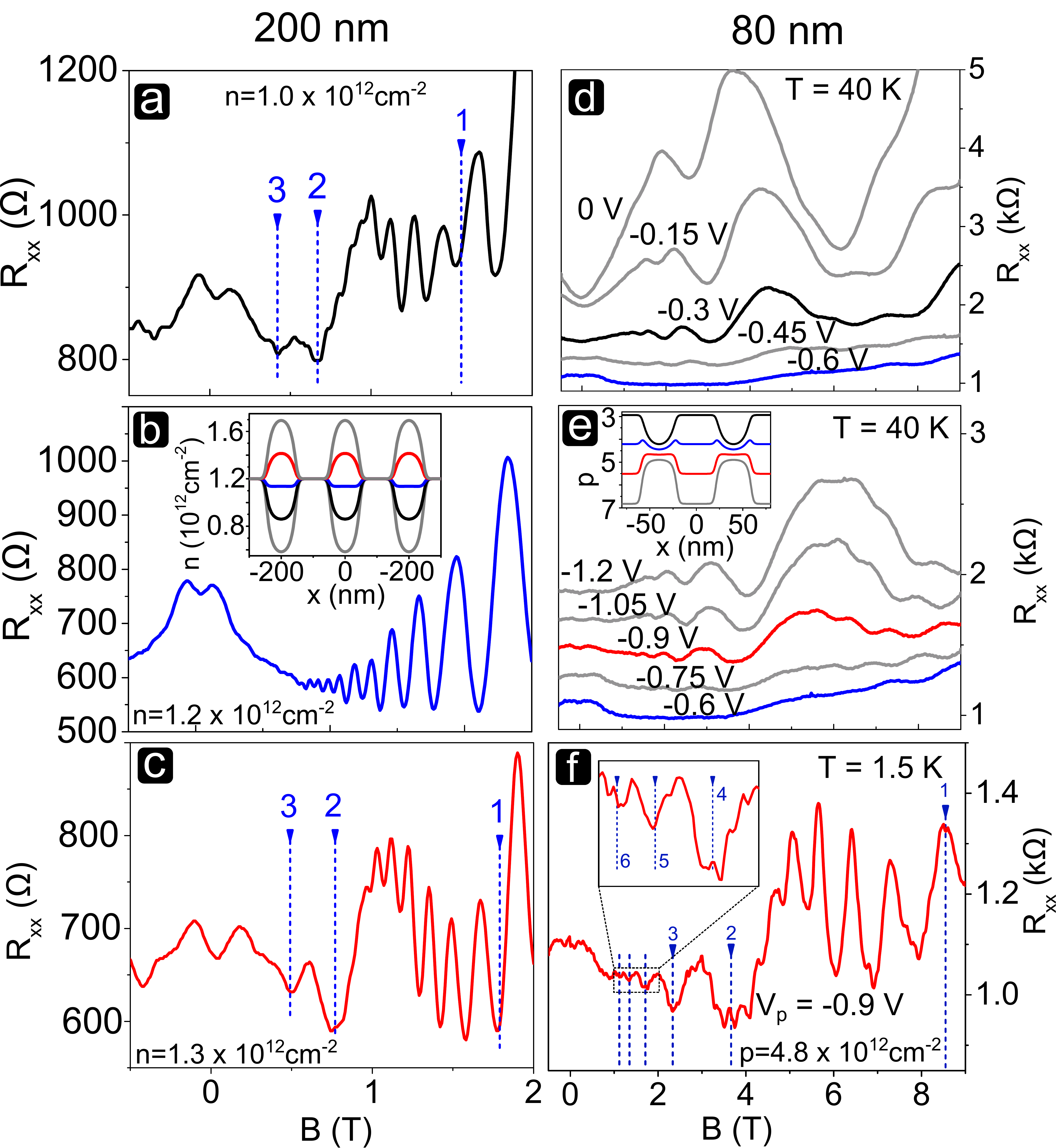}
	\caption{
		\textbf{Commensurability oscillations in graphene.} \textbf{a-c)} magnetoresistance of sample A ($a_A=200\unit{nm}$) at $T=1.5\unit{K}$ for  fixed  $V_p=0.9\unit{V}$ and $V_g=10, 15, 20\unit{V}$, respectively. The densities $n$ were extracted from SdHOs at higher fields. Vertical blue lines: calculated flat band position (Eq. \ref{eq:1}). The COs appear for weak modulation (\textbf{a,c}) and disappear in the demodulated situation (\textbf{b}). The inset in \textbf{b} shows the calculated charge carrier density profiles for blue $V_g=5...25\unit{V}$, where situations \textbf{a}-\textbf{c} are represented by colors. 
{\textbf{d,e):} Magnetoresistance of sample B ($a_B=80\unit{nm}$) at $T=40\unit{K}$, fixed $V_g=-25\unit{V}$ and 
 $V_p=0\ \dots -0.6\unit{V}$ (\textbf{d}) and  $V_p=-0.6\dots -1.2\unit{V}$ (\textbf{e})}		
 \textbf{f)}: {$T=1.5$ K,} minima up to $\lambda=6$ are resolvable. Inset in \textbf{e}: 1D charge carrier density distribution {(units of $10^{12}$ cm$^{-2}$)} for $V_p=-0.3...-1.2\unit{V}$, where gate configurations in \textbf{d,e,f} are represented by colors.}
	\label{pic:fig2}
\end{figure}

Below, we focus on the unipolar regime, to obtain a weak and tunable 1D superlattice. This is the regime of the COs outlined above. Let us first discuss magnetotransport in sample A with mobility $\mu\approx 55\,000\unit{cm^{2}/Vs}$ (see Supplemental Material \cite{Note2} for the determination of mobility in both samples). In Fig. \ref{pic:fig2}\textbf{a-c} we show three magnetic field sweeps, where we keep the PBG-voltage fixed at $V_p=0.9\unit{V}$ and tune the modulation strength by varying the backgate voltage $V_g$. The sweeps represent three different situations, \textbf{(a)} $n<n^\prime$, \textbf{(b)} $n\approx n^\prime$ and \textbf{(c)} $n>n^\prime$.
The corresponding $V_g,V_p$-positions of the sweeps \textbf{a}-\textbf{c}  are marked by stars in the $nn^\prime$-quadrant of Fig. \ref{pic:fig1}\textbf{d}. Moreover, the inset in Fig. \ref{pic:fig2}\textbf{b} shows the corresponding charge carrier density profiles that were calculated employing a 1D electrostatic model of the device, including a quantum capacitance correction \cite{drienovsky2014multibarriers, liu2013quantum_capacitance}, but neglecting screening.

In Fig.~\ref{pic:fig2}\textbf{a} a weak, unipolar ($n<n^\prime$) potential modulation is shown where the longitudinal resistance $R_{xx}$ exhibits well pronounced peaks and dips prior to the emergence of 
SdHOs, appearing at slightly higher $B$-fields. 
The average charge carrier density, extracted from SdHOs is $1.0\times10^{12}\unit{cm{}^{-2}}$ for this particular gate configuration, yielding a mean free path $l_f=\hbar k_F\mu/e=0.64\unit{\upmu m}$. The expected flat band positions (Eq. (\ref{eq:1})), are denoted by the blue vertical dotted lines, perfectly describing the experimentally observed minima. The dips are resolved up to $\lambda=3$, corresponding to a cyclotron orbit circumference of $2\pi r_C=1.7\unit{\upmu m}$. This clearly confirms that ballistic transport is maintained over {several} periods of the superlattice. 

At $V_g=15\unit{V}$ (Fig. \ref{pic:fig2}\textbf{b}), 
$n\approx n^\prime$ holds (see inset in Fig. \ref{pic:fig2}\textbf{b}). We still observe clear SdHOs, but the COs disappeared. The pronounced peak at $\sim \pm0.16\unit{T}$ can be attributed to a magneto-size effect related to boundary scattering in ballistic conductors \cite{thornton1989boundaryscattering, beenakker1991quantum} of width $W$. While in GaAs based 2DEGs, a ratio $W/r_C\approx 0.5$ is found, we extract $W/r_C \approx 1$ in accordance with previous studies on graphene \cite{masubuchi2012boundary}.

 Further increasing $V_g$ increases $n$ and switches the modulation on again ($n>n^\prime$). The SdHOs in Fig. \ref{pic:fig2}\textbf{c} yield an average $n=1.3\times 10^{12}\unit{cm^{-2}}$. Again, three minima appear at the expected flat band condition described by Eq. (\ref{eq:1}). 

Let us turn to sample B, where we demonstrate COs in the \textit{p} regime. It has a short period of $a_B=80 \unit{nm}$ and a bottom hBN flake of only $2\unit{nm}$ thickness, separating the PBG from the graphene. 
{Figure \ref{pic:fig2}\textbf{d,e} shows the longitudinal resistance at high average hole densities ($-n = p\approx 4.5\times10^{12}\unit{cm^{-2}}$)} as a function of the PBG-voltage $V_p$ and the perpendicular B-field at fixed $V_g=-25\unit{V}$ (Dirac-point at $V_g=40\unit{V}$) and $T=40\unit{K}$. Here, $T$ was increased in order to damp the SdHOs for better resolution of the COs. The mobility of {$\mu=30\,000\unit{cm^2/Vs}$} and the rather large hole density $p$ give rise to a mean free path {$l_f\approx 0.75\unit{\upmu m}$}. 
We can resolve COs up to $\lambda=6$ ({see Fig.~\ref{pic:fig2}\textbf{f}} \footnote{{Fine features in Fig.~\ref{pic:fig2}\textbf{f} are presumably due to mesoscopic fluctuations.}}), corresponding to a cyclotron orbit circumference of $2\pi r_C=1.4\unit{\upmu m}$, which is {about twice} $l_f$   in the average density range considered. 
At around $V_p\sim-0.6\unit{V}$ the COs disappear as the modulation potential becomes minimal ({blue lines in Figs. \ref{pic:fig2}\textbf{d,e}, {\em cf.} charge carrier density profile in the inset})
Here, $p\approx p^\prime$ holds and no COs are resolvable. 
This changes again at 
{$V_p=-0.6\dots -1.2\unit{V}$ (Fig. \ref{pic:fig2}\textbf{e}). As the back gate voltage is further increased, strong COs appear again, with the minima positions shifting according to the density dependence of Eq. \eqref{eq:1} (see also Supplemental Material for a color map \cite{Note2}).}
The observation of clear cut COs in density modulated hole and electron systems for distinctively different superlattice periods highlights the suitability of graphene PBGs for imposing lateral potentials on graphene films.

As pointed out in the introduction, theory predicted enhanced COs in graphene \cite{matulis2007appearance}. To check this and to compare theory and experiment we  apply  the different prevailing theoretical models to describe $R_{xx}(B)$ for our sample.
The amplitude of the COs is governed by the period $a$, the modulation amplitude $V_0$ and the Drude transport relaxation time {$\tau_{p}$}. 
Expressions for the additional band conductivity $\Delta\sigma_{yy}$ for 2DEG in \cite{peeters1992CO-Tdep} and graphene in \cite{matulis2007appearance} (Eqs. S1, S9 in Supplemental Material, respectively \cite{Note2}), are linear in $B$ and tend to overestimate the CO amplitude at lower field. Mirlin and W\"olfle \cite{mirlin1998weiss} introduced anisotropic scattering to the problem by taking into account the small angle impurity scattering, allowing for a high ratio of the momentum relaxation time {$\tau_{p}$} to the elastic scattering time $\tau_{e}$ (Eq. S10, Supplemental Material \cite{Note2}). In this approach, both the damping of COs at lower fields and the modulation amplitudes of conventional 2DEGs are correctly described. For the graphene case, Matulis and Peeters employed the Dirac-type Landau level spectrum, as opposed to the parabolic 2DEG situation \cite{matulis2007appearance}, leading to a modified expression. In their approach, only a single transport scattering time {$\tau_{p}$} was included. The temperature dependence of the COs was treated in Refs. \cite{beton1990temperature, peeters1992CO-Tdep} for parabolic 2DEGs and in Ref. \cite{matulis2007appearance} for graphene. It is expected to exhibit a 
$x/\sinh(x)$-dependence, where $x=T/T_{c}$ with the critical temperature 
 \begin{equation}
 T_{c} = \frac{Bea}{4\pi^2k_B}v_F.
 \label{eq:Tcrit}
 \end{equation}
 Here, $k_B$ is the Boltzmann constant and the difference between parabolic and linear dispersion is absorbed in the different Fermi velocities $v_F$. 

\begin{figure}[t]
	\includegraphics[width=0.95\columnwidth]{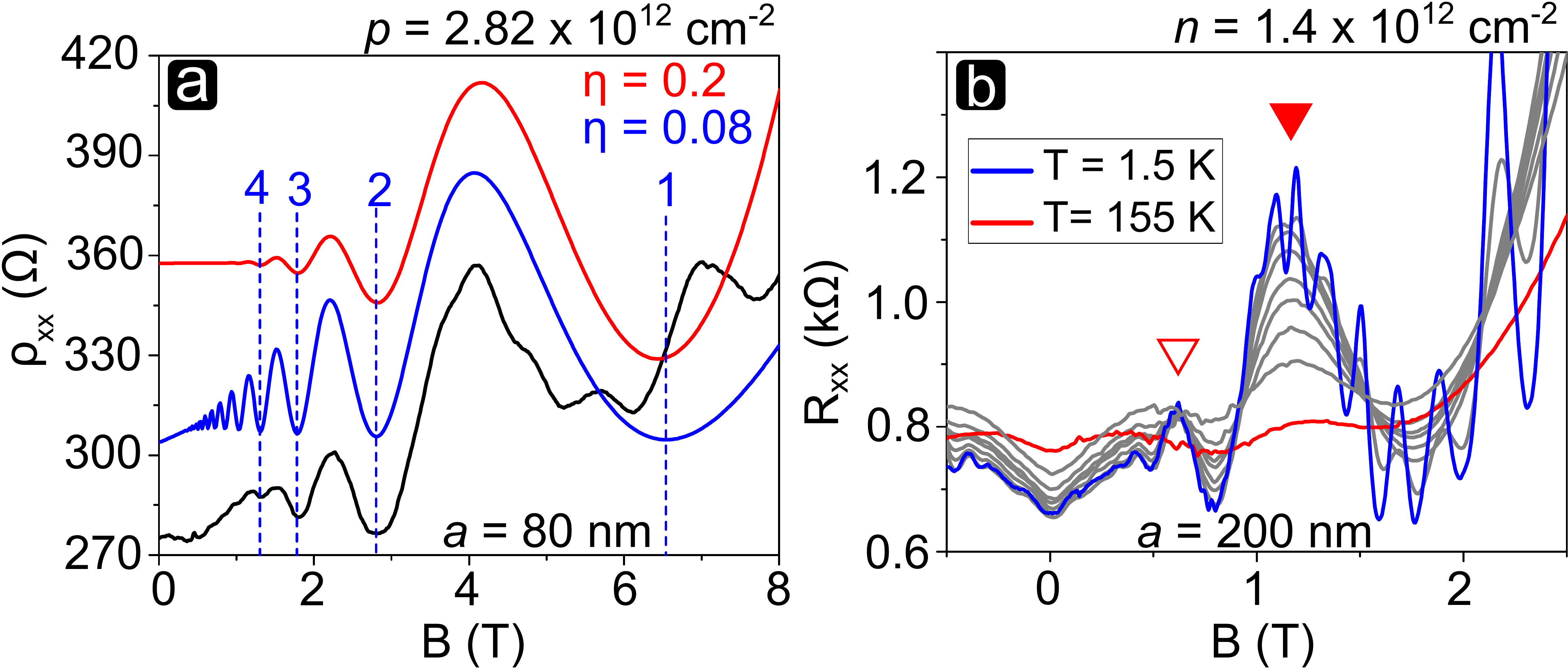}
	\caption{\textbf{a}): Comparison of different theoretical expressions with experiment at $V_g=-25\unit{V}$, $V_p=-0.8\unit{V}$ and $T=40\unit{K}$ (black curve). The blue curve represents the theory for graphene \cite{matulis2007appearance} with isotropic scattering and {$\eta=0.08$}. The red curve includes small angle impurity scattering \cite{mirlin1998weiss}, using {$\eta=0.2$}, matches the experiment well. The experimental curve does not show a pronounced minimum around 6 Tesla due to strong SdHOs setting in. {\textbf{b}): Longitudinal magnetoresistance of sample A at $V_g=25\unit{V}$ and $V_p=0.9\unit{V}$ at different temperatures, from $1.5\unit{K}$ to $155\unit{K}$.}}
	\label{pic:Fig3}
\end{figure}

 To compare to the different theoretical models, 
we extracted the elastic scattering time $\tau_{e}=(80\pm10)\unit{fs}$ from the SdHO-envelope \cite{monteverde2010transport} of a reference Hall bar (see Supplemental Material for details \cite{Note2}). With {$\mu=30\,000\unit{cm^2/Vs}$}, we obtain the ratio  {$\tau_{p}/\tau_{e}\approx 7.4$}, which emphasizes the importance of small angle scattering in hBN-encapsulated graphene.
The experimental (black) curve in Fig. \ref{pic:Fig3}\textbf{a} was taken at $T=40\unit{K}$, where the SdHOs are already visibly suppressed, but the amplitude of the COs is practically unchanged, allowing for a better comparison to theory.  We first compare our measurement to the graphene theory employing isotropic scattering only. Since the superlattice period $a_B=80\unit{nm}$, {$\tau_{p}=0.59\unit{ps}$}, temperature $T=40\unit{K}$ and the average charge carrier density $p=2.8\times10^{12}\unit{cm^{-2}}$ are known, only {the relative modulation strength} $\eta=V_0/E_F$ remains as a fitting parameter. 
By fitting the theoretical expressions (for details see Supplemental Material \cite{Note2})
to the CO peak at $\approx 4$ T in Fig. \ref{pic:Fig3}\textbf{a} we obtain {$\eta=0.08$}. 
At lower fields, the experimentally observed oscillations decay much faster than the calculated ones.
Inserting our sample parameters into the theory employing small angle impurity scattering \cite{mirlin1998weiss} (Eq. S10, Supplemental Material \cite{Note2}),
again only $\eta=V_0/E_F$ remains as a free fitting parameter, and we obtain the red trace using {$\eta=0.2$}.  
The fit describes the experimental magnetoresistance strikingly well, although the Dirac nature of the spectrum was not considered. 
The fits in Fig. \ref{pic:Fig3}\textbf{a} 
imply that including small angle impurity scattering is essential for the correct description of encapsulated graphene. 

\begin{figure}[t]
	\includegraphics[width=1.0\columnwidth]{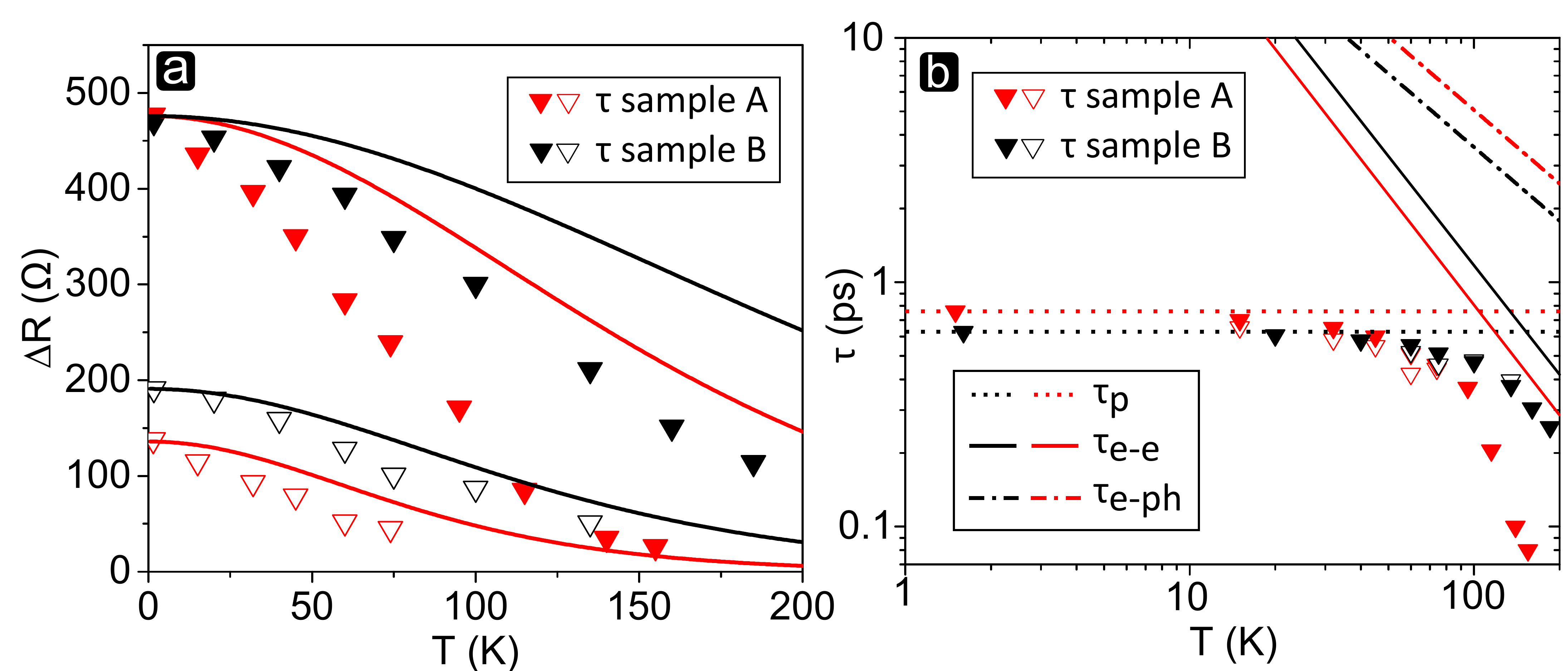}
	\caption{\textbf{a):}  T-dependence of the CO peaks, marked by the red triangles in {Fig. \ref{pic:Fig3}\textbf{b}}. Black triangles: {similar data for sample B at density $p=2.82\times10^{12}\unit{cm^{-2}}$. Solid lines: Expected $T$-dependence $x/\sinh(x)$, where $x=T/T_c$ and $T_c$ from Eq. \eqref{eq:Tcrit}.
\textbf{b):} Scattering times extracted for both samples, together with $\tau_p$ from $T=1.5$ K and predictions for $\tau_{e-e}$ \cite{kumar2017superballistic} and $\tau_{e-ph}$ \cite{hwang2008eph}.} 
	}
	\label{pic:Fig4}
\end{figure}

Finally, we discuss the temperature dependence of COs in 1D modulated graphene.  Figure \ref{pic:Fig3}\textbf{b} depicts a longitudinal resistance trace of sample A at $n=1.4\times10^{12}\unit{cm{}^2}$ at different temperatures. The graph clearly demonstrates that the COs are much more robust than the SdHOs. While the latter are almost completely suppressed at $T=40\unit{K}$, the COs survive at least up to $T=150\unit{K}$ (Sample A) and $T=200\unit{K}$ (Sample B), respectively. We analyze the temperature evolution of the first two CO-peaks (marked by red triangles), using 
the connecting line between two adjacent minima as the bottom line {to evaluate the height of the maximum in between. We adopt this procedure described by Beton {\em et al.} for a better comparison to experiments in GaAs} \cite{beton1990temperature}. 
The temperature dependence of the two peaks is shown in Fig. \ref{pic:Fig4}\textbf{a}. Also shown are the corresponding data of sample B (black symbols).
{The data for sample B were extracted at much higher fields, due to the smaller lattice period and higher carrier density, leading to a weaker temperature dependence in Eq. \eqref{eq:Tcrit}. The expected temperature dependence (solid lines) $(T/T_c)/\sinh(T/T_c)$ \cite{beton1990temperature,peeters1992CO-Tdep,matulis2007appearance}, clearly deviates from the experimental data points. For GaAs, $T$-dependent damping of the COs was so strong that the assumption of a $T$-independent scattering time was justified \cite{beton1990temperature}. In graphene, the higher $v_F$ leads to a higher $T_c$ and therefore the COs persist to higher temperatures than in GaAs. Hence, we have to consider a $T$-dependence of the scattering time as well. Using the low-temperature momentum relaxation time $\tau_p$ we first determine the modulation strength $\eta$, Then, using a fixed $\eta$, we extract the scattering time $\tau$ entering into the CO theory at elevated temperatures. The extracted times are plotted in Fig. \ref{pic:Fig4}\textbf{b} for both samples and two magnetic fields each, together with predictions for the electron-electron scattering time \cite{kumar2017superballistic} and electron-phonon scattering time \cite{hwang2008eph}. Clearly, at $T\gg 10$ K, $\tau$ deviates visibly from $\tau_p$, with $\tau_{e-e}$ being the relevant cut-off, while $\tau_{e-ph}$ is not important. This resembles the recently found observation window for hydrodynamic effects in graphene \cite{Ho2018}.}


To conclude, we present the first experimental evidence of commensurability oscillations (COs) {\cite{weiss1989magnetoresistance}} for both electrons and holes in a hBN-encapsulated monolayer graphene subject to a 1D periodic potential. This was made possible through the combined action of a nanopatterned FLG bottom gate and a global Si back gate. Our approach allows tuning both carrier density and modulation strength independently in a wide range, and on the scale of a few tens of nanometers. The minima in $R_{xx}(B)$ are well described by the flat band condition (\ref{eq:1}). The predicted strong temperature robustness of COs in graphene was qualitatively confirmed, but detailed comparison to existing theories emphasized the need for a description including anisotropic scattering of charge carriers in encapsulated graphene. {Using data at elevated temperature, we could extract the $T$-dependence of the scattering time, pointing to electron-electron scattering as the high-$T$ cutoff for the CO amplitude.}

\begin{acknowledgments}
Financial support by the Deutsche Forschungsgemeinschaft (DFG) within the programs GRK 1570 and SFB 689 (projects A7 and A8) and project Ri 681/13 ``Ballistic Graphene Devices'' and by the Taiwan Minister of Science and Technology (MOST) under Grant No.
107-2112-M-006-004-MY3 is gratefully acknowledged. Growth of hexagonal boron nitride crystals was supported by the Elemental Strategy Initiative conducted by the MEXT, Japan and JSPS KAKENHI Grant Numbers JP15K21722. We thank J. Amann, C. Baumgartner, A. T. Nguyen and J. Sahliger for their contribution towards the optimization of the fabrication procedure. 
\end{acknowledgments}


%



\section*{Supplementary material}

# Commensurability oscillations in one-dimensional graphene superlattices - Supplementary Material

Martin Drienovsky[1], Jonas Joachimsmeyer[1], Andreas Sandner[1], Ming-Hao Liu (劉明豪)[2,3], Takashi Taniguchi[4], Kenji Watanabe[4], Klaus Richter[3], Dieter Weiss[1] and Jonathan Eroms[1]

[1]*Institute of Experimental and Applied Physics, University of Regensburg, D-93040 Regensburg, Germany*
[2]*Department of Physics, National Cheng Kung University, Tainan 70101, Taiwan*
[3]*Institute of Theoretical Physics, University of Regensburg, D-93040 Regensburg, Germany*
[4]*National Institute for Materials Science, 1-1 Namiki, Tsukuba 305-0044, Japan*

## 1 Sequential etching

With the introduction of few layer graphene patterned bottom gates (PBG) [S1] to the van der Waals stacking method, the accurate control of the etch depth becomes challenging. The $CHF_3/O_2$-recipe (table S1, step 8) commonly used for etching of hBN/graphene/hBN-stacks is not appropriate to simultaneously create stable edge contacts and control the etch depth in the lower hBN (l-hBN) layer on the nanometer-scale.

To overcome this problem, we introduce a sequential etching procedure, making use of the high selectivity of $SF_6$ [S2] and $O_2$, for hBN and graphene, respectively. The sequence of processing steps is listed in table S1. The etching steps are sketched in Fig. S1. After electron beam lithography (EBL) (step 1), we remove resist residues in the e-beam-exposed areas by reactive ion etching (RIE) with $O_2$-plasma (step 2). This prevents a pronounced grass effect in the following $SF_6$-RIE (step 3). Due to its high selectivity, $SF_6$ etches hBN very fast, while PMMA and graphene serve as a hard mask. For step 3 we use a "mild" RIE-recipe with a low share of physical etching, so that the graphene, covering the l-hBN and PBG, is preserved. After removal of the PMMA with acetone (step 4), the residual graphene alongside the mesa is etched by $O_2$ (step 5). The l-hBN and hence the PBG, remain unaffected. Fig. 1**a** of the main text shows the mesa of sample B after step 5. We spin-coat the sample with PMMA (step 6) and expose the contact areas of the Hall bar and the PBG by EBL (step 7). For proper edge contacts at the voltage probes of the mesa and of the PBG, we etch the exposed contact areas with $CHF_3/O_2$ (step 8). Another $O_2$-step makes sure that no residues of the $CHF_3$-etching cover the contact areas (step 9). Finally, the contact areas are metalized by Cr(1 nm)/Au(90 nm) (step 10).

| # | type | gas | flow | p | P | t | aim/purpose |
|---|------|-----|------|---|---|---|-------------|
| 0 | spin coat | - | - | - | - | - | coat with PMMA |
| 1 | EBL | - | - | - | - | - | expose PMMA for mesa |
| 2 | RIE | $O_2$ | 20 sccm | 30 mTorr | 20 W | $30 - 90$ s | etch PMMA residues |
| 3 | RIE | $SF_6$ | 3 sccm | 50 mTorr | 10 W | $20 - 30$ s | etch hBN |
| 4 | acetone bath | - | - | - | - | - | remove PMMA |
| 5 | RIE | $O_2$ | 20 sccm | 30 mTorr | 20 W | $30 - 60$ s | remove residual graphene |
| 6 | spin coat | - | - | - | - | - | coat with PMMA |
| 7 | EBL | - | - | - | - | - | expose contact areas |
| 8 | RIE | $CHF_3/O_2$ | 40/6 sccm | 55 mTorr | 35 W | $5 - 15$ s | etch side contact areas |
| 9 | RIE | $O_2$ | 20 sccm | 30 mTorr | 20 W | 30 s | remove residues |
| 10 | metalization | - | - | - | - | - | Cr/Au side contacts |

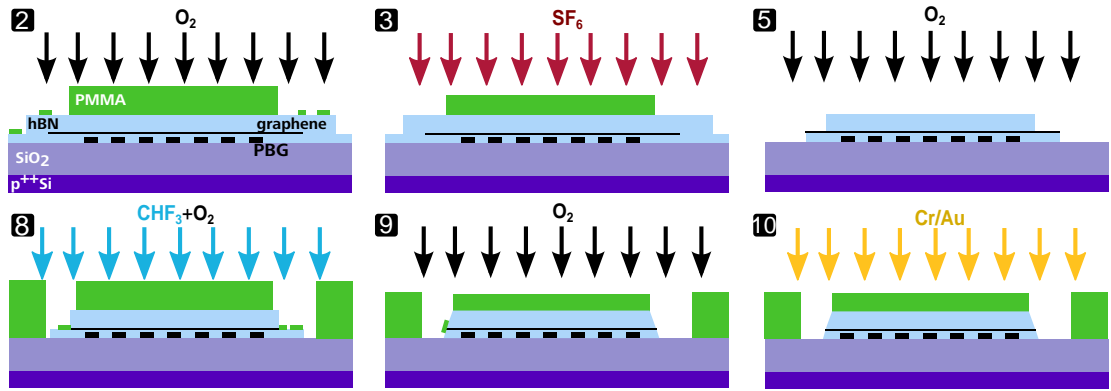

Figure S1: **Sequential etching of hBN/graphene/hBN/PBG heterostructures.** The steps are labeled as in the text or table S1, respectively. **2)** RIE with $O_2$-plasma to remove resist residues. **3)** $SF_6$ removes the exposed hBN, while PMMA and graphene serve as a mask. **5)** The exposed graphene is etched away with $O_2$. **8-9)** $CHF_3+O_2$- and $O_2$- pre-etching of contact areas. **10)** Metalization with Cr/Au.

Table S1: Sequential etching procedure for hBN/graphene/hBN/PBG on a Oxford Plasmalab 80 Plus RIE-system.

## 2 Sample B: Zero-field behavior

The zero field 4-terminal resistance of sample B is depicted in Fig. S2. Figure S2a shows the color map of the resistance during the first measurement cycle. The sample was initially *p*-doped, with the Dirac-point at $V_g \approx 10\,\text{V}$. The doping of the sample increased significantly in the time interval of a few weeks between first and second cool down (Fig. S2b) . We assume extrinsic doping of the whole sample area, since both DPs for the PBG and backgated area, respectively were shifted towards positive values. The quality of sample B however barely changed. The magnetic field behavior depicted in Fig. 2d-f of the main text was measured after the extra doping occurred, corresponding to the zero-field behavior of Fig. S2b.

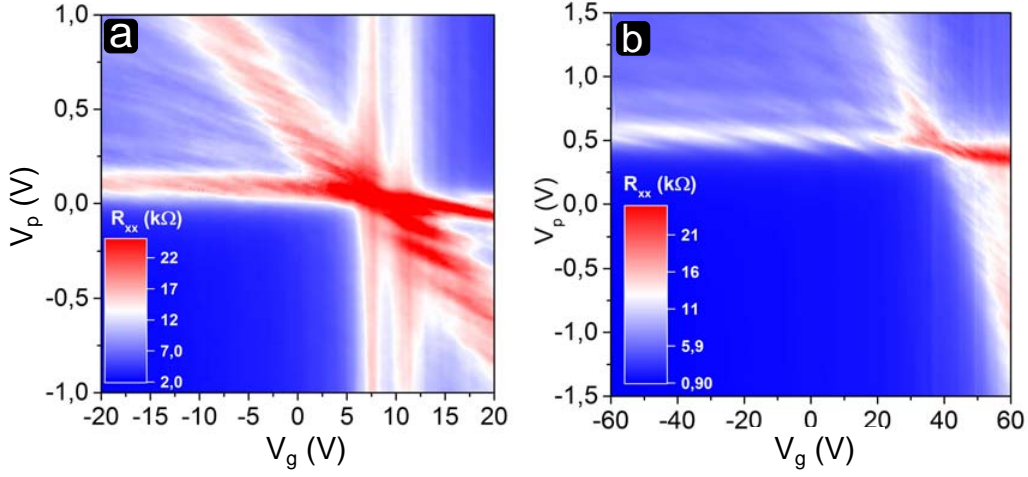

Figure S2: **Zero-field behavior of sample B. a)** Longitudinal resistance of the PBG-area after the first cool-down. **b)** During a second measurement cycle a few weeks afterwards, the *p*-doping of the sample increased.

## 3  Sample B: COs in *nn′*-regime

In the main text, we focus on commensurability oscillations (COs) of sample B in the *pp′*-transport regime. To underline the ambipolarity of graphene, we present an example of COs in the *n*-type region as well. Figure S3 shows COs at an average electron density $n = 7.3 \times 10^{11}\,\mathrm{cm^{-2}}$. The minima calculated from the flat band condition, Eq. (1) of the main text, coincide well with the experiment. The data shown in Fig. S3 was collected during the first measurement cycle, represented by Fig. S2**a**.

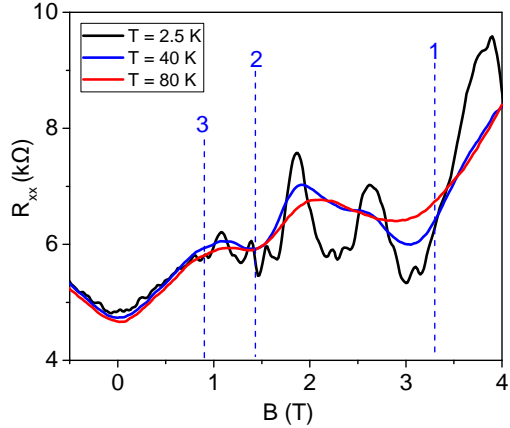

Figure S3: **COs in the *nn′*-regime of sample B.** Longitudinal resistance of sample B in a perpendicular magnetic field. At 2.5 K, SdHOs superimpose the COs. By increasing the temperature, the SdHOs vanish and the more robust COs remain, showing 3 pronounced minima, marked by calculated, dashed vertical lines. Data taken at $V_g = 25\,\mathrm{V}\ V_p = 0.2\,\mathrm{V}$.

## 4  COs in a 2DEG with isotropic scattering

A weak, periodic potential modulation in one dimension, superimposed onto a 2DEG, (i.e. a 1D-superlattice) gives rise to commensurability oscillations (COs) of the longitudinal magneto-resistivity $\rho_{xx}$ [S3]. The modulation acts as a weak perturbation to the system and the Landau levels (LLs) are

broadened into bands, with oscillating bandwidth [S4, S5]. The dispersion of these bands gives rise to an increased band conductivity $\sigma_{yy}$, perpendicular to the modulation. The flat band condition, Eq. (1) in the main text, gives the $B$-field-dependent position where the band conductivity vanishes and longitudinal resistance is minimal. The extra contribution to conductivity $\Delta\sigma_{yy}$ for a conventional 2DEG, employing an isotropic transport scattering time $\tau_{tr}$, can be expressed as [S6, S7]:

$$\Delta\sigma_{yy} = A_0 \frac{T}{8\pi^4 T_c} \left( (1-A) + 2A\cos^2\left( \frac{2\pi r_C}{a} - \frac{\pi}{4} \right) \right), \tag{S1}$$

with $r_C = \hbar k_F / eB$ the cyclotron radius, $a$ the superlattice period and the prefactor

$$A_0 = \frac{8\pi^2 e^2 V_0^2 \tau_{tr}}{h\hbar k_B T}, \tag{S2}$$

where $k_B$ is the Boltzmann-constant. The temperature dependence is given by

$$A = \frac{T}{T_c} \frac{1}{\sinh(T/T_c)}. \tag{S3}$$

The longitudinal resistivity $\rho_{xx}$ can be obtained via inversion of the conductivity tensor:

$$\begin{pmatrix} \rho_{xx} & \rho_{xy} \\ \rho_{yx} & \rho_{yy} \end{pmatrix} = \frac{1}{\sigma_{xx}\sigma_{yy} - \sigma_{yx}\sigma_{xy}} \begin{pmatrix} \sigma_{yy} & -\sigma_{xy} \\ -\sigma_{yx} & \sigma_{xx} \end{pmatrix}, \tag{S4}$$

with

$$\sigma_0 = ne\mu, \qquad \sigma_{xx} = \frac{\sigma_0}{1+\mu^2 B^2} \qquad \sigma_{xy} = \frac{\sigma_0 \mu B}{1+\mu^2 B^2}. \tag{S5}$$

$n$ is the electron density and $\mu \propto l_f = v_F \tau_{tr}$ the mean free path, with the Fermi velocity $v_F$ and the momentum relaxation time $\tau_{tr}$. For an isotropic system $\sigma_{yy}^{iso} = \sigma_{xx}$; $\sigma_{yx} = -\sigma_{xy}$. Introducing an anisotropy via the 1D-modulation, the overall conductivity, perpendicular to the modulation is given by:

$$\sigma_{yy} = \sigma_{yy}^{iso} + \Delta\sigma_{yy}. \tag{S6}$$

Hence, employing Eqs. (S1-S6), the resistivity can be expressed as:

$$\rho_{xx} = \frac{\sigma_{yy}}{\sigma_{xx}\sigma_{yy} + \sigma_{xy}^2}. \tag{S7}$$

Equation (S1) is an asymptotic expression for large quantum numbers and employed to Eq. S7, it reproduces the general behavior of the longitudinal magnetoresistance for a weakly modulated, conventional 2DEG. The thermal damping (Eq. (S3)), is governed by the critical temperature [S8]:

$$T_c = \frac{Bea}{4\pi^2 k_B} v_F. \tag{S8}$$

In the limit of isotropic scattering, the amplitude of COs increases linearly with *B* (cf. Fig. S4).

## 5  COs in graphene with isotropic scattering

For weakly modulated graphene, Matulis and Peeters employed a similar approach as for the 2DEG case and obtained for the extra band conductivity [S7]:

$$\Delta\sigma_{yy} = A_0 \frac{T}{8\pi^4 T_D} \cos^2\left(\frac{\pi}{k_F a}\right) \left((1-A) + 2A\cos^2\left(\frac{2\pi r_C}{a} - \frac{\pi}{4}\right)\right), \tag{S9}$$

with the prefactor $A_0$, given by Eq. S2. Expressions (S1) for the conventional 2DEG and (S9) for graphene are quite similar. Both systems yield the same *T*-dependence (Eqs. (S3) and (S8)). Still, a stronger robustness of COs in graphene with respect to temperature is expected [S7], as the Fermi velocity of the two systems is, in general, quite different. While the 2DEG exhibits a *k*-dependent carrier velocity, Dirac fermions in graphene move with constant $v_F = c/300$, independent of their energy [S9], where *c* is the speed of light in vacuum. The consequence of this discrepancy is stressed in Fig. S4, where the calculated $\rho_{xx}$ of a 1D-modulated, conventional GaAs-2DEG and graphene are shown. The curves are plotted for the same set of parameters $a, \mu, T$ and $\eta = V_0/E_F$. For 2DEG and graphene, the different nature of charge carriers and hence different effective mass, leads to different amplitudes of COs. At equal *n* (Fig. S4a), the CO-frequency is shifted due to the different $k(n)$-dependence of 2DEG and graphene. For equal $r_C$, the larger amplitude of COs in graphene becomes even more apparent (Fig. S4b).

The extra term $\cos^2(\pi/k_F a)$ in Eq. (S9) is a consequence of the Dirac-like spectrum of graphene giving rise to a slightly different set of Laguerre-polynomials in the analytic expression of $\Delta\sigma_{yy}$ [S7, S10]. However, its role is limited, as it is $\approx 1$ in the experimentally accessible range of charge carrier density *n* (see inset of Fig. S4).

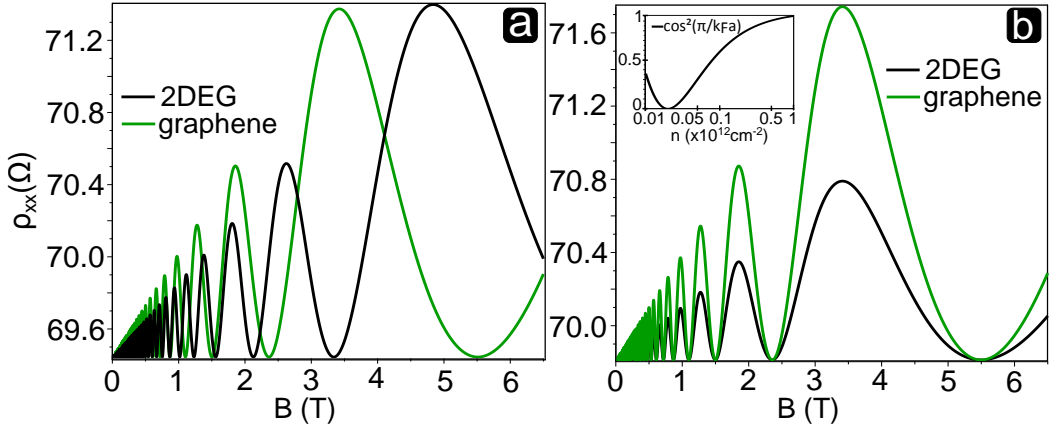

**Figure S4: COs in a 2DEG and graphene with isotropic scattering.** Magneto-resistance $\rho_{xx}$, calculated from Eqs. (S1) and (S9) for **a)** equal $n = 2 \times 10^{12}\,\mathrm{cm}^{-2}$ and **b)** equal $k_F = 2.5 \times 10^8\,\mathrm{m}^{-1}$ or $r_C B = 164\,\mathrm{nmT}$, respectively. The other parameters are $a = 80\,\mathrm{nm}$, $\eta = 0.01$, $\mu = 45\,000\,\mathrm{cm}^2/\mathrm{Vs}$ and $T = 1.5\,\mathrm{K}$. The effective mass for the parabolic ($p$) spectrum of the 2DEG was assumed $m_p^* = 0.067 m_0$. The inset shows the term $\cos(\pi/k_F a)$ from Eq. (S9) dependent on $n$.

For the curves shown in Fig. S4, the ratio of critical temperatures [S7] and hence the relative thermal robustness of the Dirac ($D$) and parabolic ($p$) system, is $T_D/T_p = v_F^D/v_F^p = 1.6$ (Fig. S4a) and $T_D/T_p = v_F^D/v_F^p = 2.3$ (Fig. S4b), respectively. Figure S5 depicts the temperature evolution of the curves shown in S4b and emphasizes that within the approximation of isotropic scattering, COs in graphene are more pronounced with respect to a 2DEG with similar properties.

## 6 COs in a 2DEG with anisotropic scattering

A 2DEG at the interface of a GaAs heterostructure is in general well separated from the surrounding and hence influenced only by long range potentials, originating from impurities far above the transport channel. Such random potential fluctuations favor small angle scattering where the momentum relaxation time $\tau_{tr}$ and the elastic (quantum) scattering time $\tau_e$ can differ by one order of magnitude. Mirlin and Wölfle [S11] introduced small angle impurity scattering to the problem of COs and found that the CO-amplitudes deviate from linear behavior. An analytical expression, covering the whole $B$-field range is given by [S11]:

$$\frac{\Delta\rho_{xx}}{\rho_0} = \frac{\eta^2}{4} Q^2 S_{tr} \frac{\pi}{\sinh \pi\zeta} J_{i\zeta}(Q) J_{-i\zeta}(Q),\qquad(\mathrm{S}10)$$

with

$$\eta = V_0/E_F, \quad Q = 2\pi r_C/a, \qquad S_{tr} = \tau_{tr}\omega_C, \qquad J_{i\zeta}(Q) \text{ the Bessel-function},\qquad(\mathrm{S}11)$$

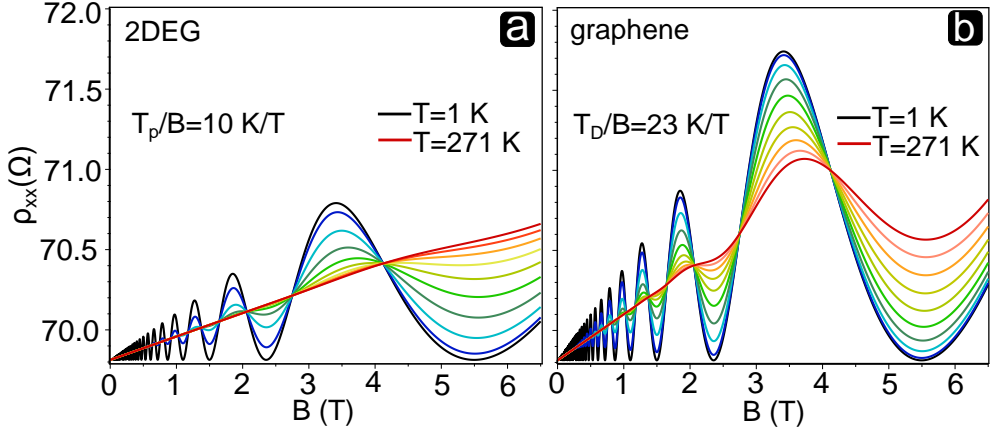

Figure S5: **$T$-dependence of COs for isotropic scattering.** The curves show calculated COs at $T = 1 \dots 271\,\mathrm{K}$ in 30 K-steps. Equation (S8) yields $T_p/B = 10\,\mathrm{K/T}$ and $T_D/B = 23\,\mathrm{K/T}$. The other parameters are $k_F = 2.5 \times 10^8\,\mathrm{m}^{-1}$, $a = 80\,\mathrm{nm}$, $\eta = 0.01$, $\mu = 45\,000\,\mathrm{cm}^2/\mathrm{Vs}$.

and

$$\zeta = \frac{1}{S_e} \left( 1 - \left( 1 + \frac{\tau_e}{\tau_{tr}} Q^2 \right)^{-1/2} \right), \tag{S12}$$

with $\tau_e$ being the scattering time of small angle scattering events and $S_e = \tau_e \omega_c$. The expression (S10) well describes the non-linear behavior of $\rho_{xx}$ in a weakly modulated 2DEG [S11], first observed by Weiss *et al.* [S3]. A plot of the function can be found in the main text, Fig. 3.

## 7 Determination of elastic scattering time

Since the samples discussed in the main text are hBN/graphene/hBN heterostructures, we expect the influence of impurities to be of similar nature as in a GaAs heterostructure. To estimate $\tau_e$, we make use of the approach presented by Monteverde *et al.* [S12], where the normalized magneto-resistance is described by the following formula:

$$\frac{\delta \rho_{xx}(B)}{\rho_0} = -4 D_T \exp\left[ -\frac{\pi}{\omega_C \tau_e} \right] \cos\left[ \frac{\pi E_F}{\hbar \omega_C} - \pi \right], \tag{S13}$$

with the temperature-dependent damping:

$$D_T = \frac{\gamma}{\sinh \gamma}, \qquad \gamma = \frac{2\pi^2 k_B T}{\hbar \omega_C}. \tag{S14}$$

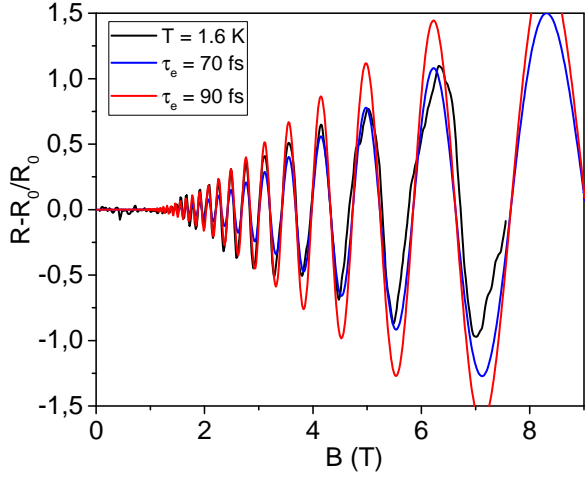

Figure S6: **SdHOs in reference area 3 of sample B.** Normalized resistivity (black) and calculated curves employing the maximum (red) and minimum (blue) estimate of $\tau_e$.

We fit Eq. (S14) to the measured magneto-resistance of the reference area 3 (see Fig. 1a in the main text). The best fits to the SdHO in low (large) field are obtained with $\tau_e = 90\,\text{ps}$ ($\tau_e = 70\,\text{ps}$) (Fig. S6).

## 8 Comparison of COs in hBN/graphene/hBN with theory

In Fig. 3 of the main text, we compare our data, measured at 40 K, to the theory, discussed in the previous sections. The period of our superlattice is well defined by lithography. In order to get a good match to the frequency of COs, we extract the charge carrier density from Shubnikov-de Haas oscillations (SdHO). Moreover, we acquire the mobility, the scattering times $\tau_{tr}$ and $\tau_e$ from measurements of an unpatterned reference area of the sample (see previous section). Hence, $V_0$ remains the only unknown parameter. We fit our curves with theoretical $\rho_{xx}$-traces, calculated from Eq. S9 and S10, where in the latter case we use $k_F = \sqrt{\pi n}$, $m^* = \hbar k_F / v_F$, $\omega_C = eB/m^*$ and $r_C = \hbar k_F / eB$, representing graphene. The experimental curve is matched by compensating the offset of $\rho_0$ and fitting the amplitude of the first CO-peak, that is in Fig. 3 at $B \approx 4\,\text{T}$.

## 9 Determination of mobility

For sample A, no reference section was available, so we determined the mobility in the demodulated regime at $n = 1.84 \times 10^{12}\,\text{cm}^{-2}$ and $T = 1.5\,\text{K}$. The carrier density was extracted from the period of SdH-oscillations, and the resistivity was taken outside the enhanced resistivity due to boundary scattering around $B = 0$. For sample B, a reference section without patterned bottom gate was available on the same graphene flake. Again, the resistivity was taken outside the enhanced resistivity around $B = 0$.

## 10 Additional magnetotransport data

In the demodulated regime of sample A, we observe a well-defined quantum Hall effect reflected in a Landau fan plotted in Fig. S8. For this, both back gate and PBG were changed simultaneously,

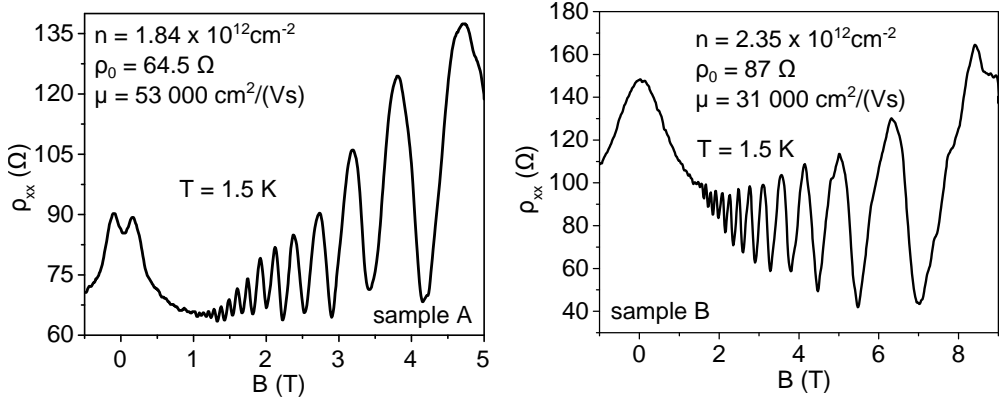

Figure S7: Determination of mobility. Left: Sample A in the demodulated regime. Right: Reference section of sample B.

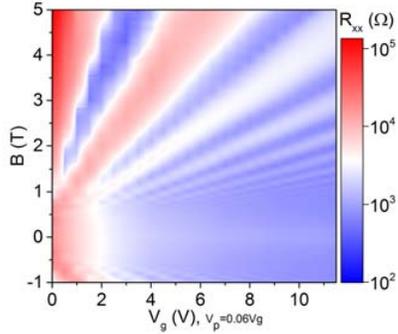

Figure S8: Landau fan diagram for $n \approx n'$ of sample A, where $V_p = 0.06 V_g$ was tracking the dashed line in Fig. 1d of the main text.

following the dashed line in Fig 1d of the main text.

The color map in Fig. S9(a) shows the $B$-field response at high average hole densities ($-n = p \approx 4.5 \times 10^{12} \, \mathrm{cm}^{-2}$) at $T = 40$ K, to suppress SdHOs. The white dotted lines serve as a guide to the eye for the flatband positions, showing that the positions of the CO minima shift with increasing carrier density, as expected. The white dashed line at $V_p = -0.6$ V corresponds to minimal modulation. A corresponding magnetoresistance curve at $T = 1.5$ K is shown in Fig. S9(b). Note that due to the thin lower hBN dielectric, the modulation can never be turned off completely, leading to suppressed visibility of SdHOs, unlike in sample A, where a completely demodulated regime can be obtained. The white dashed line at $V_p = -0.9$ V corresponds to the low temperature plot shown in Fig 2f of the main text.

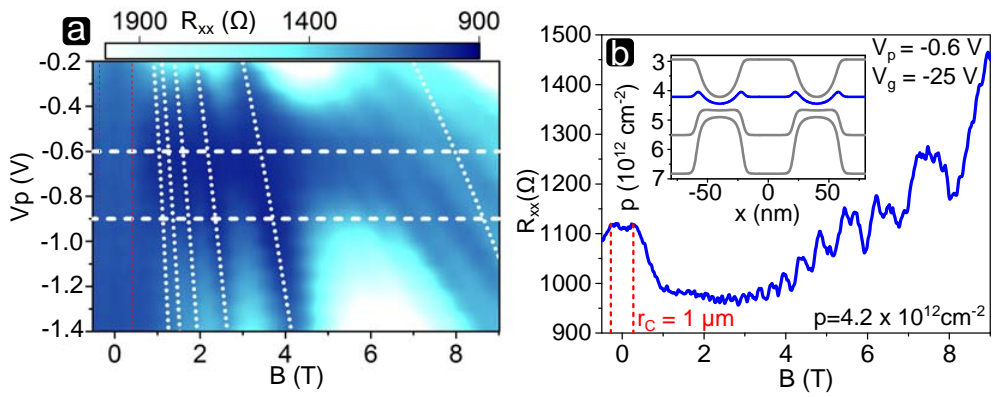

Figure S9: Additional magnetotransport data of sample B. (a): color plot showing the disappearance and reappearance of COs at $T = 40$ K, as the PBG voltage is ramped from -0.2 V to -1.4 V. (b): Magnetoresistance at $T = 1.5$ K, with the modulation reduced to the minimum. The calculated density profile (blue line in the inset) shows that a residual modulation always remains. Red dashed lines: position of ballistic feature due to boundary scattering at the edges of the narrow Hall bar.



\end{document}